\documentstyle[12pt]{article}
\date{ }
\title{Trace anomaly of the conformal gauge field}

\author{J. S\l adkowski $^{\dag \$ }$  \\
Fakult\" at f\" ur   Physik Universt\" at Bielefeld,\\
D-4800 Bielefeld 1, Universitatsstrasse 25, Germany}
\begin{document}
\baselineskip9mm
\maketitle
\begin{abstract}
\baselineskip9mm
   The proposed by Bastianelli and van Nieuwenhuizen new method of
calculations of trace anomalies is applied in the  conformal
gauge field case. The result is then reproduced by the heat
equation method. An error in previous calculation is corrected.
It is pointed out that the introducing gauge symmetries into
a given system by a field-enlarging transformation can result in
unexpected quantum effects even for trivial configurations.

\end{abstract}
\vspace{15mm}

$^{\dag}$ A. von Humboldt Fellow; permanent address: Dept.
of Field Theory and
Particle Physics, University of Silesia, Pl 40007 Katowice,
Poland.\\

$^{\$}$ E-mail: sladk@usctoux1.cto.us.edu.pl and sladk@hrz.uni-bielefeld.de

\newpage

\subsection*{\ \ 1\ Introduction}

The trace anomaly in a scalar theory
with possible additional fields, although
less mundane than its chiral counterpart [1, 2],
has been intensively studied for a
long time. This is because in quantum field theory on curved
spacetime, the conformal symmetry is an important issue.
One can obtain a conformally invariant Lagrangian from
 the Lagrangian

$$ {\it L}= \frac{1}{2}\int d^{n}x \sqrt{-g} g^{\mu \nu}
\partial _{\mu} \phi \partial _{\nu}\phi \eqno(1)$$
by adding the term $-\zeta R\phi^{2}$, where R denotes the scalar
curvature and $\zeta =\frac{n-2}{4\left( n-1\right)}$. The resulting
Lagrangian is then invariant with respect to:
$$g_{\mu \nu}\	\rightarrow \ g'_{\mu \nu}= e^{2\alpha}
g_{\mu \nu} \ ;\ \
\phi \ \rightarrow \ \phi'= e^{\left( 2-n\right)
\alpha} \phi , \eqno(2)$$
where $\alpha$ is an arbitrary real function. It is also possible
to obtain a Lagrangian invariant with respect to (2) by
introducing an additional
vector field to (1) [3]. The appropriate Lagrangian

$$ {\it L}= \frac{1}{2}\int d^{n}x \sqrt{-g} g^{\mu \nu}\left(
\partial _{\mu} + A_{\mu}\right) \phi \left( \partial _{\nu} +
A_{\nu} \right) \phi \eqno(3)$$
is then invariant with respect to (2) provided the vector field
transforms as

$$A_{\mu} \ \rightarrow \  A'_{\mu} = A_{\mu} +
\partial _{\mu} \alpha , \ \ \  n>2 \eqno(4)$$
that is as a gauge field (conformal gauge field). For $n=2$ the
gauge field $ A_{\mu}$ is invariant (and redundant).
The gauge group is the multiplicative group
$R - \{ 0 \}$, where $R$ denotes the real numbers. What
more interesting is one can combine both methods of achieving
conformal invariance in (1). The result is

$$ {\it L}= \frac{1}{2}\int d^{n}x \sqrt{-g} \{ g^{\mu \nu}\left(
\partial _{\mu} + A_{\mu}\right) \phi \left( \partial _{\nu} +
A_{\nu} \right) \phi - a\phi ^{2}\left( \zeta R + A_{\mu}A^{\mu} +
\nabla _{\mu} A^{\mu} \right) \} \ ,  \eqno(5)$$
where an additional parameter a has been introduced. The case of
$a=1$ corresponds to the ordinary conformally invariant
scalar field Lagrangian.

\subsection*{\ \ 2\ The trace anomaly}

The conformal
gauge field is not a gauge field in the usual sense. An
ordinary gauge field is conformally invariant. Note that for
$n\not= 4$ one cannot add the conformal gauge field kinetic
term $F_{\mu \nu}F^{\mu \nu}$ to (5) because it spoils the
conformal invariance. Further we will restrict ourselves
to the four dimensional case. It seems to be interesting to apply
the proposed recently by Bastianelli and Nieuwenhuizen method of
calculating the trace anomaly [4]. This is a generalization of the
ideas of Alvarez-Gaume' and Witten [5] to the conformal case.
The method consists in replacing n-dimensional operators like
$x^{\mu},\ \partial _{\mu},\ \gamma ^{\mu} \
g_{\mu \nu},\  A^{\mu}, \ and \ T^{a}$
(generator of the gauge group) by the
their quantum mechanical analog. All you have to worry about
is to keep the appropriate (anti)commutation rules. One
then gets the following formula for the
conformal anomaly (the reader is referred to [4] for details):

$$ I^{n}= \lim_{\beta \to 0} 'Tr\ e^{-\beta H}=
\lim_{\beta \to 0} ' \frac{1}
{\left( 2\pi\beta\right)^{n\over{2}}}\int d^{n}x\sqrt{g}<0\vert
Te^{- {{1}\over{\beta}}S_{int}} \vert 0> \ , \eqno(6) $$
where the prime over the lim symbol denotes that we should
take only the $\beta$ independent (finite) part .
One can directly follow the rout proposed in [4]. The
Hamiltonian that corresponds to (5) have the form:

$$ H= - {{1}\over{2}} \left( \nabla^{\mu} + A^{\mu} \right)
\left( \partial _{\mu} + A_{\mu} \right) - {1\over 8}R + V \ ,
\eqno(7)$$
where the "scalar potential" V is given by
$$V= {a\over 2}\left( {1\over 6}R
+ A^{\mu}A_{\mu} + \nabla ^{\mu}A_{\mu} \right) + {1\over 8}R$$.
By repeating the calculation and using the proposed in [4] time
ordering prescriptions one gets

$$\begin{array}{lll}
I^{4} & = & {-1\over {28800\pi ^{2}}}\{ 10R_{\mu \nu\sigma\tau}R^
{\mu \nu\sigma\tau}
- 10R_{\mu \nu}R^{\mu \nu} -60\Box R + 25 R^{2} + 150F_{\mu \nu}
F^{\mu \nu}\cr
\ &\ &\ \cr
\  & \	& + 300a \left( \Box - R\right) \left( {1\over 6}R +
A^{\mu}A_{\mu}
+\nabla ^{\mu} A_{\mu}\right) + 900a^{2}\left( {1\over 6}R +
A^{\mu}A_{\mu}
+\nabla ^{\mu} A_{\mu}\right)^{2}\}\cr
\end{array}  \ . \eqno(8)$$
This result is in apparent disagreement with the one presented
in [6, 7]. To check our result we will repeat the calculation of
the trace anomaly by the heat kernel method [8]. The
trace anomaly in four dimensional spacetime is given by [9]

$$I^{4}= -\frac{a_{2}}{16\pi ^{2}} \ , \eqno(9)$$
where $a_{2}$ is calculated for the operator
$$D= -g^{\mu \nu}\left( \nabla^{\mu} + A^{\mu} \right)
\left( \partial _{\mu} + A_{\mu} \right) + V  \ , \eqno(10)$$
where
$$V=a\left( {1\over 6}R
+ A^{\mu}A_{\mu} + \nabla ^{\mu}A_{\mu} \right) \ . \eqno(11)$$
The calculation is straightforward and leads to

$$\begin{array}{lll}
I^{4} & = & {-1\over {28800\pi ^{2}}}\{ 10R_{\mu \nu\sigma\tau}R^
{\mu \nu\sigma\tau}
- 10R_{\mu \nu}R^{\mu \nu} -60\Box R + 25 R^{2} + 150F_{\mu \nu}
F^{\mu \nu}\cr
\ &\ &\ \cr
\ & \  &  + 300 \left( \Box - R\right) V + 900V^{2} \}\cr
\end{array}  \ . \eqno(12)$$
By inserting the V given by (11) we reproduce the formula (8).
So the the method of [4 ] gives the correct result. One can can
also calculate the trace in a more general case, e.g. when
fermions are present. You have only to worry about the
appropriate form of the Hamiltonian and the time ordering
prescriptions.\\

\subsection*{\ \ 3\ Concluding remarks}

\ \ \ We would like to conclude by several remarks. First, the use
of the "Feynman approach" (perturbative expansion) gives another
interpretation in terms interaction vortices of various terms
that appear in the trace formula. The reader is encouraged to
repeat at least part of the calculation presented in [4].
Secondly, the the perturbative expansion (6) can be compared with
the formula [9]

$$I^{n}= -\frac{ a_{n\over 2}}{\left( 4\pi \right)^{n\over 2}}
\eqno(13)$$
and used to analyse subtleties of the $\zeta$-function
regularization method [10-12].
For example it can be used to determine the coefficients
$a_{n}\left( x, x'\right)$ in the limit $x \to x'$ for n even:

$$ a_{n}= finite\ part\ of\ \lim_{\beta \to 0}
\frac{-2^{n\over 2}}{\beta ^{n\over 2}}<0\vert	Te^{-
{{1}\over{\beta}}S_{int}} \vert 0> \eqno(14) \ . $$

Finally, let us notice that the Lagrangian (3) can be also
obtained by the field-enlarging transformation [13, 14]:
$$g_{\mu \nu} \rightarrow g_{\mu \nu}
e^{2\alpha}\eqno(15)$$

and
$$ \phi \rightarrow \phi e^{-\alpha} \ . \eqno(16)  $$
Then, if one introduces
the conformal gauge field $A_{\mu}$ via $ A_{\mu}=
\partial _{\mu} \alpha $,
one gets (3). Of course, in this case $F_{\mu \nu} = 0$ so that
the kinetic term is absent from (3) and (5). Even such a
trivial conformal gauge configuration can produce a non-vanishing
anomaly. This points out that the introducing gauge symmetries
into a given model
by a field-enlarging transformations can
result in unexpected quantum effects because it may not
be possible to find a local interaction term that cancel the
anomaly contribution. One can use a local scale transformation
to remove the scalar field (set $\phi =1 $). Then one gets
an Einstein gravity coupled to an auxilliary vector field.
Such a model is not scale-invariant. We can say that it is a
conformal gauge fixed form of (5).

{\bf Acknowledgements}. The author would like to
thank prof. R. K\"  ogerler and dr K. Ko\l odziej for
stimulating and helpful discussions. This work has been
supported in part by the Alexander von Humboldt Foundation and
the Polish Committee for Scientific Research under the contract
KBN-PB 829/P3/92/03.

\newpage
\subsection*{\ \ References}

\newcounter{bban}

\begin{list}
{[\arabic{bban}]}{\usecounter{bban}\setlength{\rightmargin}
{\leftmargin}}

\item J. S\l adkowski and M. Zra\l ek, Phys. Rev. {\bf D45}
(1992) 1701.
\item M. Abdelhafiz, K. Ko\l odziej and M. Zra\l ek,
Jagellonian Univ.
preprint TPJU-2/86 (1986).
\item T. Padmanbhan, Class. Quantum Grav. {\bf 2} (1985) L105.
\item F. Bastianelli and P. van Nieuwenhuizen, Nucl. Phys. {\bf
B389} (1993) 53; F. Bastianelli,
Nucl. Phys. {\bf B376} (1992) 113.
\item L. Alvarez-Gaume' and E. Witten, Nucl. Phys. {\bf B234}
(1984) 269.
\item M. S. Alvez and J. Barcelos-Neto, Class. Quantum Grav.
{\bf 5} (1988) 377.
\item M. S. Alvez and C. Farina, Class. Quantum Grav. {\bf 9}
 (1992) 1841.
\item B. DeWitt, Dynamical theory of groups and fields
(Gordon and Breach, New York, USA, 1965).
\item N. D. Birrel and P. C. W. Davies, Quantum fields in curved
spacetime (Cambridge University Press, Cambridge, U. K. 1984).
\item S. Hawking, Commun. Math. Phys. {\bf 55} (1997) 133.
\item R. Ma\'nka and J. S\l adkowski, Phys. Lett. {\bf B224}
(1989) 97.
\item R. Ma\'nka, Class. Quantum Grav. {\bf 4} (1987) 1789.
\item J. S\l adkowski, Phys. Lett. {\bf B296} (1992) 361.
\item J. Alfaro and P. H. Damgaard, Ann. Phys (NY) {\bf 202}
(1990) 398.

\end{list}

\end{document}